\documentstyle[psfig]{mn}

\title[Mrk~1014: An AGN Dominated ULIRG at X-rays]
{Mrk 1014: An AGN Dominated Ultraluminous Infrared Galaxy
}

\author[Th. Boller et al.]
{Th. Boller, L.C. Gallo, D. Lutz and E. Sturm\\
Max-Planck-Institut f\"ur extraterrestrische Physik,
Postfach 1603, 85748 Garching, Germany \\
\\
}
\date{Received 2002 May 30; Accepted 2002 July 17 }

\begin{document}
\label{firstpage}

\maketitle

\begin{abstract}
In this paper we report on an \em XMM--Newton\em~observation of the ultraluminous infrared QSO
Mrk~1014.  The X-ray observation 
reveals a power-law dominated ($\Gamma \approx$ 2.2) spectrum with a
slight excess in the soft energy range.  
AGN and starburst emission models fit the soft excess emission equally well, however,
the most plausible explanation is an AGN component as the 
starburst model parameter, temperature and luminosity, appear physically unrealistic.
The mean luminosity of Mrk~1014 is about 2 $\times$ 10$^{44}$ erg s$^{-1}$. 
We have also observed excess emission at energies greater than 5 keV.  This feature could be
attributed to a broadened and redshifted iron complex, but deeper observations are required to
constrain its origin.  The light curve shows small scale
variability over the $\sim$11 ks observation.  There is no evidence of intrinsic absorption
in Mrk~1014.  The X-ray observations support the notion of an AGN dominated central engine.
We establish the
need for a longer observation to constrain more precisely the nature of the X-ray components.
\end{abstract}


\section{Introduction}
Mrk 1014 ($V = 15.7$; $z = 0.163$),
with a far-infrared (FIR) luminosity in excess of 10$^{12}$L$_{\odot}$ (Yun 2001),
is one of the brightest members of the class of `warm' ultraluminous
infrared galaxies (Sanders et al. 1988a) which
from their warm IRAS 25$\mu$m/60$\mu$m colors and optical spectra,
are believed to host powerful AGN. 
Sanders et al. (1988b) report on the detection of CO(1--0) emission and suggest that 
Mrk~1014 may be an important link in the evolution of ultraluminous infrared galaxies 
into UV-excess quasars.

Spectroscopy using the
\em Infrared Space Observatory (ISO)\em~adds quantitative information to this
picture. The low resolution ISOPHOT-S spectrum presented by Rigopoulou et al.
(1999) is dominated by AGN-like continuum emission, with no evidence for
the aromatic `PAH' emission typical for starbursts. The ratio of
7.7$\mu$m PAH feature and local continuum is $<$0.63, well below starburst
values of $\sim$3. The diagnostics proposed by Genzel et al.
(1998) and Tran et al. (2001) then suggest an upper limit of $\approx$25\%
for the starburst contribution to the infrared luminosity of Mrk~1014.
Sturm et al. (2002) present ISO-SWS mid-IR fine structure line data
for Mrk~1014. The source is at the limit of the the SWS sensitivity,
and only tentative detections are obtained for the low excitation
[Ne\,II]12.8$\mu$m line and the high excitation [O\,IV]25.9$\mu$m line,
with a ratio consistent with AGN dominance.

\em Hubble Space Telescope (HST)\em~NICMOS observations (Scoville et al. 2000) reveal 
twisting spiral 
isophotes beneath the dominant QSO nucleus, 
indicating either a starburst spiral disk or tidal debris.
Optical observations of the host galaxy portray a prominent tidal arm
with several knots (Surace \& Sanders 2000; Surace et al. 1998).
The tidal arm is composed primarily of intermediate age stars 
($\sim$ 1 Gyr) with very little 
contribution from older stars (Canalizo \& Stockton 2000).
Nolan et al. (2001) determine that the majority of the flux is
associated with an instantaneous burst of star formation at approximately
12~Gyr, however, Canalizo \& Stockton (2000) find several regions of 
recent ($\sim$ 0.2 Gyr) star formation accounting for up to 30\% of the total
luminous mass along the line of sight. 
Both age estimations are too high to be associated with on-going
starburst activity ($\sim$10 Myr), which produces X-ray emitting hot
gas.

Mrk~1014 was observed with \em XMM--Newton\em~(Jansen et al. 2001) as part of the guaranteed
time programme. The objective of the study was to obtain a high quality 
X--ray spectrum of an AGN dominated ultraluminous infrared galaxy, and
determine if starburst activity is a necessary component to explain the spectral
energy distribution. 

In the following section we will present the observations and discuss the data reduction.
In section 3, we will describe the spectral models and discuss their significance.
A description of the light curve will be presented in section 4. 
In section 5 we will summarize our findings. 
A value of the Hubble constant of $H_0$=$\rm 70\ km\ s^{-1}\ Mpc^{-1}$
and a cosmological deceleration parameter of $q_0 = \rm \frac{1}{2}$ have
been adopted throughout.

\section{OBSERVATION AND DATA REDUCTION}
Mrk~1014 was observed with \em XMM--Newton\em~on 2000 July 29 during
revolution 0117 for a duration of 15 ks.
All instruments were functioning normally during this time.
The European Photon Imaging Camera (EPIC) pn detector (Str\"uder et al. 2001)
was operated in full--frame mode, and the two MOS cameras (Turner et al. 2001) 
were operated in large--window mode.  All three EPIC CCD cameras used the medium
filter.

The MOS and pn Observation Data Files (ODFs) were processed to produce
calibrated event lists using the \em XMM-Newton\em~Science Analysis System (SAS) v5.3.
Unwanted hot, dead or flickering pixels were removed as were events due to
electronic noise.  Event energies were corrected for charge--transfer losses.
The latest available calibration 
files\footnote{m1\_medv9q20t5r6\_all\_15.rsp and m2\_medv9q20t5r6\_all\_15.rsp
for MOS1 and MOS2, respectively and
epn\_ff20\_sdY9\_medium.rmf for the pn.} were used in the
processing.  Single and double events were selected.  Background light curves 
were extracted from these event lists
to search for periods of flaring, but showed the background to be stable
throughout the duration of the observations.
The total amount of ``good'' exposure time was 10534 s with the MOS cameras
and 11298 s with the pn camera.

The source plus background photons were extracted from a circular region
of radius 40 arcsec  for the pn and MOS data.
The centroid position of Mrk~1014 in the EPIC images is
$\alpha$=01$^h$59$^m$50$^s$.2,
$\delta$=+00$^{\circ}$23$^{\prime}$41.2$^{\prime\prime}$, which is in good agreement with the
optical position given by Argyle \& Eldridge (1990).
Background events were extracted from an off--source circular region roughly 90 arcsec in radius.

The Galactic column density toward Mrk~1014 is (2.6 $\pm$ 0.2) $\times$
10$^{20}$ cm$^{-2}$ (Dickey \& Lockman 1990).

\section{SPECTRAL ANALYSIS}
The source spectra were grouped such that each spectral bin contains at
least 60 counts for the pn data and 20 counts for the MOS.  
Spectral fitting was accomplished with the XSPEC v11.1
software package (Arnaud 1996).
The quoted errors on the derived best--fitting model parameters correspond
to a 90\% confidence level unless stated otherwise.  When we calculate
luminosities, we shall assume isotropic emission unless stated otherwise.

Displayed in Figure~1 is the EPIC pn spectrum along with the background spectrum
in the 0.3--10 keV energy range.  The background is no longer negligible 
at energies higher than about 8 keV.  The same is true for the MOS observations.
We will only consider the 0.3--8.0 keV spectral range when generating models of 
Mrk~1014.
\begin{figure}
\psfig{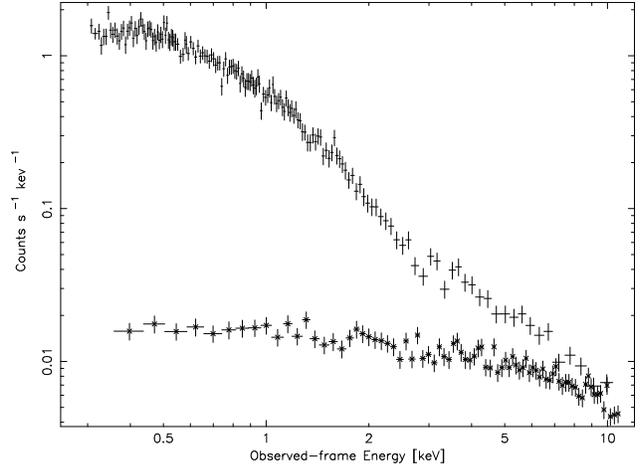}
\caption{The dominant spectrum is the source+background observed with
the EPIC pn camera.  The weaker spectrum, averaging approximately 0.01
counts s$^{-1}$ keV$^{-1}$, is the background contribution.  The
background spectrum becomes just as significant as the source spectrum
at energies higher than 8 keV.  The same is true for the MOS observations.
}
\end{figure}
In the initial models the photoelectric absorption
was treated as a free parameter.  Absorption above the Galactic $N_H$ was not
found in any spectral model indicating very little intrinsic absorption in Mrk~1014. 
In the remainder of the analysis we have fixed the absorption parameter to the Galactic
value stated above.

A simple power--law modified by Galactic absorption was
fitted separately to the pn and combined MOS 0.3--8 keV spectra with mediocre 
results ($\chi^2$ = 160.4/159 $dof$ for the
pn data, and $\chi^2$ = 252.6/229 $dof$ for the combined MOS data).  As we will demonstrate
in section 3.1, the addition of a soft X-ray component significantly improves the 
fits.  The photon
indices were $\Gamma$ = 2.46 $\pm$ 0.03 and $\Gamma$ = 2.42 $\pm$ 0.04
for the pn and MOS data, respectively.  
Despite the reasonable agreement between the pn and MOS data
we do not combine the spectra in order that we may search for discrepancies in 
the more complicated models discussed below.  We will alternate our discussion between the two
spectra and will explicitly state any incongruity between models. 

There is no convincing evidence of a narrow iron K$\alpha$ emission in either spectra
nor is there evidence of spectral variability.
A hardness ratio light curve was produced using the
0.3--2~keV and 2--8 keV energy bands to test for spectral variability,
but the hardness ratio light curve is represented by a constant 
fit ($\chi^2$ = 23.5/55 $dof$).

\subsection{The Soft Energy Spectrum}
The statistics indicate that the fits are
significantly improved by the addition of a soft component to the spectral models. 
Given the nature of Mrk~1014, this soft component may be attributed either to AGN or possible
starburst activity in the nucleus.  We put forth a number of models to test
both hypotheses.  A simple blackbody, double blackbody and multiple blackbody
disc models are used to search for an AGN contribution; thermal bremsstrahlung and a 
Raymond \& Smith (1977) models are suggested to investigate the starburst nature.
In the remainder of this section we will discuss these models.  We have summarized
the results in Table 1.

As can be seen in Figure~2, a blackbody plus power-law provides an acceptable fit to 
the data ($\chi^2$ = 145.7/156 $dof$ for the pn data, and $\chi^2$ = 210.7/212 $dof$ for
the combined MOS data).  The derived power-law indices are similar in the pn model
($\Gamma$ = 2.28 $\pm$ 0.06) and the MOS model ($\Gamma$ = 2.21 $\pm$ 0.08).  The
unabsorbed luminosities for the model components are also consistent: We measure 
a total 0.3--8 keV luminosity of
1.65 $\times$ 10$^{44}$ erg s$^{-1}$ with MOS, of which the blackbody component accounts for 16\%.
With the pn model we calculate a luminosity of 1.58 $\times$ 10$^{44}$ erg s$^{-1}$ , of which 13\% 
is attributed to the blackbody component.  A double blackbody plus power-law also provided
a good fit for the pn ($\chi^2$ = 139.5/154 $dof$) and MOS ($\chi^2$ = 209.9/210 $dof$) data, however
several parameters were unconstrained, such as the temperature of the second blackbody. This 
would indicate that the model is not an accurate representation of the data given the available 
photon statistics.
A multiple blackbody disc in addition to the power-law
was also a reasonable fit to the data ($\chi^2$ = 142.2/156 $dof$ and 210.3/212 $dof$ for the pn and MOS,
respectively).
\begin{figure}
\psfig{figure=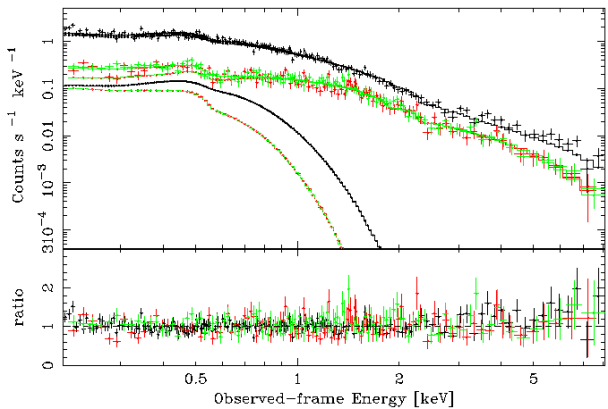,width=8.3cm,clip=}
\caption{The 0.3--8 keV EPIC pn and combined MOS spectra of Mrk~1014 and model components
are displayed in the upper panel.  The data are
fitted with a blackbody plus power-law components ($\chi^2$ = 145.7/156 $dof$ and 210.7/212 $dof$ for
for the pn and MOS, respectively).  The solid line is the pn model and the broken
line is the combined MOS fit.  The lower panel show the residuals (data/model) of
the models.  Notice the excess residuals at energies higher than
5 keV.
}
\end{figure}

In an attempt to model a possible starburst component we employed the Raymond-Smith
model for an emission spectrum from a hot, diffuse gas.  In combination with a simple
power-law the models fit the data well ($\chi^2$ = 152.2/156 $dof$ for the pn data  
and $\chi^2$ = 215.8/212 $dof$ for the MOS
data).  
We find $kT$ = 273$^{+100}_{-50}$ eV, which is lower than the temperature
of $>$600 eV usually found in X-ray starbursts (Ptak et al. 1999).
This claim is further supported by the recent \em XMM-Newton\em~observations of
the ULIRG NGC~6240.  The starburst in NGC 6240 is adequately fitted by two Raymond-Smith
components with temperatures $kT$ = (0.62 $\pm$ 0.20) keV and (1.11 $\pm$ 0.10) keV (Keil,
Boller \& Fujimoto, in preparation).
We calculate an X-ray luminosity of 
$\approx$ 5 $\times$ 10$^{42}$ erg s$^{-1}$ for the Raymond-Smith component in Mrk~1014.
This X-ray luminosity appears to be surprisingly large for a starburst,
therefore, the detection of a starburst component above the luminous AGN power-law 
is very unlikely.
A second respectable fit was obtained by adding a thermal bremsstrahlung component
to the power-law ($\chi^2$ = 139.1/156 $dof$, 210.1/212 $dof$ for the pn and MOS, respectively),
however the resulting temperatures and luminosities are unphysically large.

On a statistical bases, none of the attempted fits appear superior over the
others.  
However, from physical arguments the AGN dominated spectral model appear the most
realistic explanation for the X-ray features of Mrk 1014.
In Table 1 we present the parameters
of the various models we attempted for the pn spectrum.  Results are comparable for the MOS 
observations.  In Table 2 we have calculated the unabsorbed luminosities measured for the soft
and power-law component for each of the pn models used.  Again, the results are comparable
with the MOS observations.

The spectral analysis reveals that Mrk~1014 is AGN dominated.  This is further supported
by comparing the soft X-ray (0.1--2.4 keV) flux to the FIR (40--120 $\mu$m) flux.
Boller \& Bertoldi (1996) found that in equilibrium, the ratio of the
X-ray to FIR flux is about 10$^{-2.5}$.
A flux ratio less than 10$^{-2.5}$ indicates that starburst
activity is dominant.  The total FIR flux of Mrk~1014
is 9.93 $\times$ 10$^{-11}$ erg s$^{-1}$ cm$^{-2}$, computed from
the IRAS 60 and 100 $\mu$m fluxes following Helou et al. (1985).  To compute
the soft X-ray fluxes we used the models described in section 3.
The extrapolated 0.1--2.4 keV \em XMM-Newton\em~flux is
3.68 $\times$ 10$^{-12}$ erg s$^{-1}$ cm$^{-2}$.
We calculated a value of 10$^{-1.4}$ for the flux ratio in Mrk~1014, an order of
magnitude greater than what is expected from a starburst.

\begin{table}
\caption[]{
Spectral Parameters of the Attempted EPIC pn Models  
}
\label{tab1}
\begin{flushleft}
\begin{tabular}{cccc}\hline\hline
Soft Energy Model & $kT$ (eV) & Power-law Index ($\Gamma$) & $\chi^2_{\nu}$ \\ \hline
Blackbody  & 147 $\pm$ 15 & 2.3 $\pm$ 0.1 & 0.93       \\
Double Blackbody$^{1}$  & 74$^{+27}_{-20}$  & 2.1$^{+0.2}_{-0.1}$ & 0.91     \\
Mult. Blackbody Disc$^{2}$ & 179$^{+18}_{-9}$ & 2.2 $\pm$ 0.1 & 0.91       \\
Raymond-Smith & 273$^{+100}_{-50}$  & 2.4 $\pm$ 0.4  & 0.98 \\   \hline\hline 
\end{tabular}
{$^1$}{Data is presented for only one of the blackbodies as the
second blackbody was not constrained.} \\
{$^2$}{The temperature presented is that at the inner disc radius.}
\end{flushleft}
\end{table}

\begin{table}
\caption[]{
The 0.3--8 keV Unabsorbed Luminosity of the Various Model Components
($\times$10$^{44}$ erg s$^{-1}$)}
\label{tab2}
\begin{flushleft}
\begin{tabular}{cccc}\hline\hline
Soft Energy Model & $L_{soft}$ & $L_{power-law}$ & $\frac{L_{soft}}{L_{total}}$ \\ \hline
Blackbody  & 0.20 & 1.38  & 0.13       \\
Double Blackbody  & 0.45 & 1.16 & 0.28               \\
Mult. Blackbody Disc  & 0.30  & 1.29 & 0.19       \\
Raymond-Smith & 0.05 & 1.55 & 0.03 \\  \hline\hline
\end{tabular}
\end{flushleft}
\end{table}

\subsection{The Hard Energy Excess}
As can be seen in Figure 2, there appears to be a gentle uprising in the residuals at
energies higher than 5 keV.  This trend is apparent in all of the models presented. 
By adding a second power-law with a negative slope to the models we find an improvement in the fits. 
Using the $F$-test for the addition of one free parameter, we find that we have a
significant improvement ($>$99.95 \%) in the quality of the fits.
The current observation does not allow us to constrain the nature of this excess
high energy emission.  It would be especially interesting to determine if this hard excess
can be attributed to a broadened and redshifted iron complex.  Longer observations are required
to further investigate the characteristics of this hard energy excess. 

\section{TIMING ANALYSIS}
In Figure 3 we present the 0.3--8 keV EPIC pn light curve in 400~s bins.
The MOS light curves are similar to the pn light curve, but
only the pn data will be presented here given the higher signal to noise.
\begin{figure}
       \psfig{figure=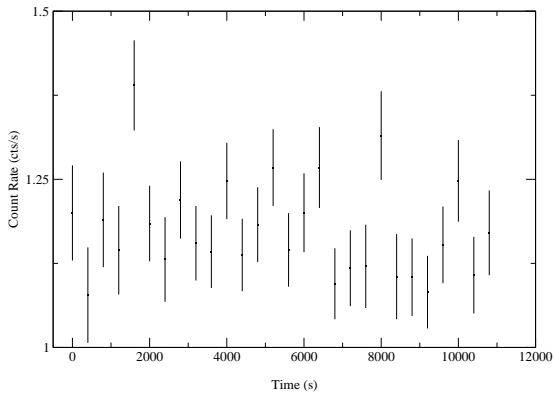,angle=-90,width=8.3cm,clip=}
      \caption{The EPIC pn light curve of Mrk 1014 in the 0.3--8 keV
energy range with a bin size of 400~s. The ordinate is the count
rate displayed in count~s$^{-1}$, and the abscissa is the time offset from
the start of the exposure on 2000 July 29 at 21:29 UT. 
}
\end{figure}
 
The light curve of Mrk~1014 does not show much in the way of interesting features but there does
appear to be some small scale variability during the 
short 11 ks observation.  A constant model through the mean of the light curve is not an acceptable 
fit ($\chi^2$ = 39.1/27 $dof$).
The mean pn count rate during the observation is (1.18 $\pm$ 0.09)
counts s$^{-1}$.  There is a difference of less than 0.3 counts s$^{-1}$ between the strongest
and weakest events in the pn light curve. The largest fluctuations in the light curve are observed
at $\sim$1600 s and $\sim$8000 s, in which the count rate increases by 0.2 counts s$^{-1}$ in
400 s.  The limited statistics in these peaks do not allow us to ascertain more information. 
The mean count rate measured with the MOS data is {1.17 $\pm$ 0.07} counts s$^{-1}$.


\section{SUMMARY}
The 0.3--8 keV spectrum of the Mrk~1014 was fitted with a number of models
to determine if the soft energy emission above a predominant power-law is 
driven by an AGN central engine or starburst activity. 
The \em XMM-Newton\em~observation reveals a power-law dominated spectrum
which may be up to thirty times more luminous than the soft X-ray excess. 
The existence of an X-ray starburst is highly unlikely due to the low temperature and
high luminosity found in the models.
Other evidence, such as a variable light curve
and the soft X-ray to FIR flux ratio suggest an AGN powered continuum
in Mrk~1014.
A longer observation with \em XMM-Newton\em~would allow us to better constrain
the nature of the soft energy component.  In addition, a longer look will
allow us to probe the interesting high energy excess above 5 keV and determine 
if it is connected to a broadened and redshifted iron complex.
Clearly, more multi-frequency observations are required 
to disentangle and compare the starburst and 
AGN contributions in different energy bands. 

\section*{Acknowledgments}
We thank the anonymous referee for constructive comments and suggestions
for improvements.
This research has made use of the NASA/IPAC Extragalactic Database (NED) 
which is operated by the Jet Propulsion Laboratory, California Institute of
Technology, under contract with the
National Aeronautics and Space Administration
This paper is based on observations obtained with \em XMM-Newton\em, an ESA
science mission with instruments and contributions directly funded by ESA member
states and the USA (NASA).

{}

\end{document}